# A smoothing model for sample disclosure risk estimation

Yosef Rinott[1],[*] and Natalie Shlomo[2],[†]

*Hebrew University and Southampton University*

**Abstract:** When a sample frequency table is published, *disclosure risk* arises when some individuals can be identified on the basis of their values in certain attributes in the table called *key variables*, and then their values in other attributes may be inferred, and their privacy is violated.

On the basis of the sample to be released, and possibly some partial knowledge of the whole population, an agency which considers releasing the sample, has to estimate the disclosure risk.

Risk arises from non-empty sample cells which represent small population cells and from population uniques in particular. Therefore risk estimation requires assessing how many of the relevant population cells are likely to be small. Various methods have been proposed for this task, and we present a method in which estimation of a population cell frequency is based on smoothing using a local neighborhood of this cell, that is, cells having similar or close values in all attributes.

We provide some preliminary results and experiments with this method. Comparisons are made to two other methods: **1**. a *log-linear models* approach in which inference on a given cell is based on a "neighborhood" of cells determined by the log-linear model. Such neighborhoods have one or some common attributes with the cell in question, but some other attributes may differ significantly. **2** The *Argus* method in which inference on a given cell is based only on the sample frequency in the specific cell, on the sample design and on some known marginal distributions of the population, without learning from any type of "neighborhood" of the given cell, nor from any model which uses the structure of the table.

## 1. Introduction

When a microdata sample file is released by an agency, directly identifying variables, such as name, address, etc., are always deleted, variable values are often grouped (e.g., Age-Groups instead of precise age), and the data is given in the form of a frequency table. However *disclosure risk* may still exist, that is, some individuals in the file may be identified by their combination of values in the variables appearing in the data.

Samples often contain information on certain variables on which the agency's information for the whole population is limited, such as expenditure on specific items in a Household Expenditure Survey, or detailed information on variables such as children's extra curricular activities in the Social Survey of the Israel Central Bureau of Statistics.

[*]Research supported by the Israel Science Foundation (grant No. 473/04).
[†]Research supported in part by the Israel Science Foundation (grant No. 473/04).
[1]Department of Statistics, Hebrew University, Jerusalem, Israel, e-mail: rinott@huji.ac.il
[2]Department of Statistics, Hebrew University of Jerusalem, Southampton Statistical Sciences Research Institute, University of Southampton, United Kingdom, e-mail: N.Shlomo@soton.ac.uk
*AMS 2000 subject classifications:* primary 62H17; secondary 62-07.
*Keywords and phrases:* sample uniques, neighborhoods, microdata.

Often agencies have to assess the disclosure risk involved in the release of sample data in the form of a frequency table when the corresponding population table may be unknown, or only partially known. Risk arises from cells in which both sample and population frequencies are small, allowing an intruder who has the sample data and access to some information on the population, and in particular on individuals of interest, to identify such individuals in the sample with high probability. Thus, the disclosure risk depends both on the given sample, and the population. In this paper we are concerned with the issue of estimating disclosure risk involved in releasing a sample on the basis of the sample alone, assuming the population is unknown.

Let $\mathbf{f} = \{f_k\}$ denote an $m$-way frequency table, which is a sample from a population table $\mathbf{F} = \{F_k\}$, where $k = (k_1, \ldots, k_m)$ indicates a cell and $f_k$ and $F_k$ denote the frequency in the sample and population cell $k$, respectively. Formally, the sample and population sizes in our models are random and their expectations are denoted by $n$ and $N$ respectively, and the number of cells by $K$. We can either assume that $n$ and $N$ are known, or that they are estimated by their natural estimators: the actual sample and population sizes, assumed to be known. In the sequel when we write $n$ of $N$ we formally refer to expectations.

If the $m$ attributes in the table can be considered *key variables*, that is, variables which are to some extent accessible to the public or to potential intruders, then disclosure risk arises from cells in which both $f_k$ and $F_k$ are positive and small, and in particular when $f_k = F_k = 1$ (sample and population *uniques*). Suppose an intruder locates a sample unique in cell $k$, say, and is aware of the fact that the combination of values $k = (k_1, \ldots, k_m)$ happens to be unique or rare in the population. If this combination matches an individual of interest to the intruder then identification can be made with high probability on the basis of the $m$ attributes. If the sample contains information on the values of other attributes, then these can now be inferred for the individual in question, and his privacy is violated. In many countries this would constitute a violation of law. For example The Central Bureau of Statistics in Israel operates under the Statistics Ordinance (1972) which says "No information... shall be so [published] as to enable the identification of the person to whom it relates".

A *global risk measure* quantifies an aspect of the total risk in the file by aggregating risk over the individual cells. For simplicity we shall focus here only on two global measures, which are based on sample uniques:

$$\tau_1 = \sum_k \mathbb{I}(f_k = 1, F_k = 1), \qquad \tau_2 = \sum_k \mathbb{I}(f_k = 1)\frac{1}{F_k},$$

where $\mathbb{I}$ denotes the indicator function. Note that $\tau_1$ counts the number of *sample uniques* which are also *population uniques*, and $\tau_2$ is the expected number of correct guesses if each sample unique is matched to a randomly chosen individual from the same population cell. These measures are somewhat arbitrary, and one could consider measures which reflect matching of individuals that are not sample uniques, possibly with some restrictions on cell sizes. Also, it may make sense to normalize these measures by some measure of the total size of the table, by the number of sample uniques, or by some measure of the information value of the data.

Various individual and global risk measures have been proposed in the literature, see e.g., Benedetti et al. [1, 2], Skinner and Holmes [12], Elamir and Skinner [6], Rinott [8].

In Section 3 we propose and explain a new method of estimation of quantities like $\tau_1$ and $\tau_2$, using a standard Poisson model, and *local smoothing* of frequency tables. The method is based on the idea that one can learn about a given population cell from neighboring cells, if a suitable definition of closeness is possible, without relying on complex modeling. In Sections 2.1 and 2.2 we briefly describe two known methods of estimation of quantities like $\tau_1$ and $\tau_2$, and in Section 4 we provide real data experiments which compare the methods discussed.

We consider the case that **f** is known, and **F** is an unknown parameter (on which there may be some partial information) and the quantities $\tau_1$ and $\tau_2$ should be estimated. Note that they are not proper parameters, since they involve both the sample **f** and the parameter **F**.

The methods discussed in this paper consist of modeling the conditional distribution of **F**|**f**, estimating parameters in this distribution and then using estimates of the form

$$(1) \quad \hat{\tau}_1 = \sum_k \mathbb{I}(f_k = 1)\hat{P}(F_k = 1|f_k = 1), \qquad \hat{\tau}_2 = \sum_k \mathbb{I}(f_k = 1)\hat{E}[\frac{1}{F_k}|f_k = 1],$$

where $\hat{P}$ and $\hat{E}$ denote estimates of the relevant conditional probability and expectation. For a general theory of estimates of this type see Zhang [14] and references therein. Some direct variance estimates appear in Rinott [8].

## 2. Models

For completeness we briefly introduce the Poisson and Negative Binomial models. More details can be found, for example, in Bethlehem et al. [3], Cameron and Trivedi [4], Rinott [8].

A common assumption in the frequency table literature is $F_k \sim \text{Poisson}(N\gamma_k)$, independently, where $N$ is assumed to be a known parameter, and $\sum \gamma_k = 1$. Binomial (or Poisson) sampling from $F_k$ means that $f_k|F_k \sim Bin(F_k, \pi_k)$, where each $\pi_k$ is a known constant which is part of the sampling design, called the sampling fraction in cell $k$. By standard calculations we then have

$$(2) \quad f_k \sim \text{Poisson}(N\gamma_k\pi_k) \text{ and } F_k \,|\, f_k \,\sim\, f_k + \text{Poisson}(N\gamma_k(1-\pi_k)),$$

leading to the Poisson model of subsection 2.1 below.

Under this model the population size is random with expectation $N$, and so is the sample size, with expectation $N\sum_k \gamma_k \pi_k$ which we denote by $n$. In practice we have in mind that $N$ and $n$ could be estimated by the actual population and sample sizes, and these estimates could be "plugged in" where needed.

If one adds the Bayesian assumption $\gamma_k \sim \text{Gamma}(\alpha, \beta)$ independently, with $\alpha\beta = 1/K$ to ensure that $E\sum \gamma_k = 1$, then $f_k \sim NB(\alpha, p_k = \frac{1}{1+N\pi_k\beta})$, the Negative Binomial distribution defined for any $\alpha > 0$ by

$$P(f_k = x) = \frac{\Gamma(x+\alpha)}{\Gamma(x+1)\Gamma(\alpha)}(1-p_k)^x p_k^\alpha, \quad x = 0, 1, 2, \ldots,$$

which for a natural $\alpha$ counts the number of *failures* until $\alpha$ successes occur in independent Bernoulli trials with probability of success $p_k$. Further calculations yield $F_k \,|\, f_k \sim f_k + NB(\alpha + f_k, \frac{N\pi_k + 1/\beta}{N+1/\beta})$, $(F_k \geq f_k)$. Note that in this model the population size is again random with expectation $N$, and now the sample size has expectation $N\sum_k \pi_k/K$ which we denote again by $n$.

As $\alpha \to 0$ (and hence $\beta \to \infty$) we obtain $F_k \,|\, f_k \sim f_k + NB(f_k, \pi_k)$, which is exactly the Negative Binomial assumption in Section 2.2 below. As $\alpha \to \infty$ the Poisson model of Section 2.1 is obtained, and in this sense the Negative Binomial with parameter $\alpha$ subsumes both models.

Next we discuss two methods which have received much attention. They have been applied in some bureaus of statistics recently, and are being tested by others.

## 2.1. The Poisson log-linear method

Skinner and Holmes [12] and Elamir and Skinner [6] proposed and studied the following approach. Assuming a fixed sampling fraction, that is, $\pi_k = \pi$, the first part of (2) implies $f_k \sim \text{Poisson}(n\gamma_k)$, where $n = N\pi$. Using the sample $\{f_k\}$ one can fit a log-linear model using standard programs, and obtain estimates $\{\hat{\gamma}_k\}$ of the parameters. Goodness of fit measures for selecting models having good risk estimates were studied in Skinner and Shlomo [11].

Using the second part of (2) it is easy to compute *individual risk measures* for cell $k$, defined by

$$
\begin{aligned}
P(F_k = 1 | f_k = 1) &= e^{-N\gamma_k(1-\pi_k)}, \\
E[\frac{1}{F_k} | f_k = 1] &= \frac{1}{N\gamma_k(1-\pi_k)}[1 - e^{-N\gamma_k(1-\pi_k)}].
\end{aligned}
\tag{3}
$$

Plugging $\hat{\gamma}_k$ for $\gamma_k$ in (3) leads to the desired estimates $\hat{P}(F_k = 1 | f_k = 1)$ and $\hat{E}[\frac{1}{F_k} | f_k = 1]$ and then to $\hat{\tau}_1$ and $\hat{\tau}_2$ of (1).

For each $k$ we therefore obtain estimates of $P(F_k = 1 | f_k = 1)$ and $E[\frac{1}{F_k} | f_k = 1]$ which depend on $\hat{\gamma}_k$, which in turn depends on the frequencies in other cells. For example, in a log-linear model of independence, $\hat{\gamma}_k$ depends on the frequencies in all cells which have a common attribute with $k$. Thus cells that are rather different in nature, having values which are very different from those of cell $k$ in most of the attributes, influence the estimates of the parameter $\gamma_k$ pertaining to this cell.

The main goal of this paper is to study the possibility of estimating $\gamma_k$ using cells in more local "neighborhoods," having attribute values which are closer to those of the cell $k$ in cases where closeness can be defined.

## 2.2. The Argus method

This method, proposed by Benedetti et al. [1, 2], was originally oriented towards individual risk estimation, but was subsequently also applied to global risk measures, see, e.g, Polettini and Seri [7], and Rinott [8]. Argus has recently been implemented in some European statistical bureaus.

In the Argus model it is assumed that $F_k | f_k \sim f_k + NB(f_k, \pi_k)$ with an implicit assumption of independence between cells. Since $\pi_k$ are assumed known we could now calculate $P_{\pi_k}(F_k = 1 | f_k = 1)$ and $E_{\pi_k}[\frac{1}{F_k} | f_k = 1]$. However because of non response, sampling biases and errors, Argus does not use the known $\pi_k$, but rather estimates them from the *sampling weights* as discussed next.

At statistics bureaus, each statistical unit responding to a sample survey is assigned a sampling weight. This weight $w_i$ is an inflating factor that informs on the *number of units in the population that are represented by sample unit $i$*, to be used for inference from the sample to the population. It is calculated by the inverse sampling fraction that is adjusted for non-response or other biases that may occur in the sampling process. These adjustments are often carried out within post-strata (weighting classes) defined by known marginal distributions of the populations, such as Age, Sex and Geographical Location. The inverse sampling fractions are calibrated so that the weighted sample count in each post-strata is equal to the known population total; this calibration reduces under or over representation of the chosen strata due to any bias, or sampling errors.

The Argus method provides *initial estimates* of the population cell sizes of the form $\hat{F}_k = \sum_{i \in \text{cell } k} w_i$, where $w_i$ denotes the sampling weight of individual $i$ described above (see also example below).

Here is a simple example:

Suppose for simplicity that the sampling weights are based only on the sampling design, and on post stratification by a single variable, say Sex, and that the sample is designed to be a random subset consisting of one percent of the population and therefore we have the same sampling fraction of $\pi = 1/100$ in each Sex group. If males, say, have a non-response rate of 20%, and females of 0%, then the sampling weight for women in the sample would be $w_i = 100$, and for men $w_i = 100/0.8 = 125$.

If in the sample table there is a cell $k = (k_1, k_2)$ where $k_1$ stands for Male, and $k_2$ stands for the level in another attribute, such as Income, and $f_k = 20$, then in this cell all $w_i$ are 125, and $\hat{F}_k = 20 * 125 = 2500$.

Now suppose Sex is **not** one of the variables in the table to be released, but the agency knows it for all individuals in the sample. Suppose the variables in the table are Income and Occupation, and suppose now $k = (k_1, k_2)$, where $k_1$ stands for a given Income group, and $k_2$ for a given Occupation. Suppose $f_k = 20$, meaning that in the sample there are 20 individuals with the given income group and occupation, and suppose that there are 10 males and 10 females in this group. The weight $w_i = 100$ for the 10 females, and 125 for the 10 males, and therefore $\hat{F}_k = 10 * 100 + 10 * 125 = 2250$.

In the above example sampling weights reflect non response. In principle a bureau may arrive at such weights also because in the original sampling design men are under represented, or because it finds out that this is the case after post stratifying on Sex and observing that males are under represented due to some reasons (some bias, including non-response, or sampling error).

Returning to Argus, recall its initial estimates of the population cell sizes $\hat{F}_k = \sum_{i \in \text{cell } k} w_i$. Using the relation $E_{\pi_k}[F_k | f_k] = f_k / \pi_k$, the parameters $\pi_k$ are estimated using the moment-type estimate $\hat{\pi}_k = f_k / \hat{F}_k$. Note that if $F_k$ were known, this would be the usual estimate of the binomial sampling probability.

Straightforward calculations with the Negative Binomial distribution show

$$P_{\hat{\pi}_k}(F_k = 1 | f_k = 1) = \hat{\pi}_k \quad \text{and} \quad E_{\hat{\pi}_k}[\frac{1}{F_k} | f_k = 1] = -\frac{\hat{\pi}_k}{1 - \hat{\pi}_k} \log(\hat{\pi}_k).$$

Plugging these estimates for $\hat{P}$ and $\hat{E}$ in (1) we obtain the estimates $\hat{\tau}_1$ and $\hat{\tau}_2$ of the global risk measures. Note that in this method the cells are treated completely independently, each cell at a time, and the structure of the table, or relations between different cells play no role. Moreover, since this method does not involve a model which reduces the number of parameters, it is required to estimate essentially $K$ parameters, which is typically hard in sparse tables of the kind we have in mind.

## 3. Smoothing polynomials and local neighborhoods

The estimation question here is essentially the following: given, say, a sample unique, how likely is it to be also a population unique, or arise from a small population cell.

If a sample unique is found in a part of the sample table where neighboring cells (by some reasonable metric, to be discussed later) are small or empty, then it seems reasonable to believe that it is more likely to have arisen from a small population cell. This motivates our attempt to study local neighborhoods, and compare the results to the type of model-driven neighborhood as the log-linear method, and the Argus method which uses no neighborhoods.

Consider frequency tables in which some of the attributes are ordinal, and define closeness between categories of an attribute in terms of the order, or more generally, suppose that for a certain attribute one can say that some values of the attribute are closer to a given value than others. For example, Age and Years of Education are ordinal attributes, and naturally the age of 5 is closer to 6 than to 7 or 17, say, while Occupation is not ordinal, but one can try to define reasonable notions of closeness between different occupations.

Classical log-linear models do not take such closeness into account, and therefore, when such models are used for individual cell parameter estimation, the estimates involve data in cells which may be rather remote from the estimated cell.

On the other hand, as mentioned above, the *Argus* method bases its estimation only on the sampling weight of the estimated population cell. There is no learning from other cells, the structure of the table plays no role, and each cell's parameter is estimated separately.

We now describe our proposed approach which consists of using local neighborhoods of the estimated cell.

Returning to (2) we assume that $f_k \sim \text{Poisson}(\lambda_k = N\gamma_k\pi_k)$. Apart from constants, the sample log-likelihood is $\sum_{k=1}^{K}[f_k \log \lambda_k - \lambda_k]$. However if we use a model for $\lambda_k$ which is valid only in some neighborhood $M$ of a given cell, we shall consider the log-likelihood of the data in this neighborhood, that is

$$(4) \qquad \sum_{k \in M}[f_k \log \lambda_k - \lambda_k].$$

For convenience of notation we now assume that $m = 2$, that is, we consider two-way tables; the extension to any $m$ is straightforward. Following Simonoff [10], see also references therein, we use a local smoothing polynomial model.

For each fixed $k = (k_1, k_2)$ separately, we write the model below for $\lambda_{k'}$ in terms of the parameters $\boldsymbol{\alpha} = (\beta_0, \beta_1, \gamma_1, \ldots, \beta_t, \gamma_t)$, with $k' = (k'_1, k'_2)$ varying in some neighborhood of $k$:

$$\begin{aligned}
(5) \qquad \log \lambda_{k'}(\boldsymbol{\alpha}) &\equiv \log \lambda_{(k'_1,k'_2)} \\
&= \beta_0 + \beta_1(k'_1 - k_1) + \gamma_1(k'_2 - k_2) + \cdots \\
&\quad + \beta_t(k'_1 - k_1)^t + \gamma_t(k'_2 - k_2)^t,
\end{aligned}$$

for some natural number $t$. One can hope that such a polynomial model is valid with a suitable $t$ for $k' = (k'_1, k'_2)$ in some neighborhood $M$ of $k = (k_1, k_2)$. Substituting (5) into (4) we maximize the concave function

$$(6) \quad L(\boldsymbol{\alpha}) = L(\beta_0, \beta_1, \gamma_1, \ldots, \beta_t, \gamma_t) = \sum_{(k'_1,k'_2) \in M} [f_{(k'_1,k'_2)} \log \lambda_{(k'_1,k'_2)} - \lambda_{(q,r)}]$$

with respect to the coefficients in $\boldsymbol{\alpha}$ of the regression model (5). With $\arg\max L(\boldsymbol{\alpha}) = \hat{\boldsymbol{\alpha}}$, and $\hat{\beta}_0$ denoting its first component, we finally obtain our estimate of $\lambda_k = \lambda_{(k_1,k_2)}$ in the form

$$(7) \qquad \hat{\lambda}_k \equiv \lambda_k(\hat{\boldsymbol{\alpha}}) = \exp(\hat{\beta}_0),$$

where the second equality is explained by taking $k' = k = (k_1, k_2)$ in (5). The maximization by the Newton-Raphson method is rather straightforward and fast.

Each of the estimates $\hat{\lambda}_k$ requires a separate maximization as above which leads to a value $\hat{\boldsymbol{\alpha}}$ that depends on $k = (k_1, k_2)$, and a set of estimates $\lambda_{k'}(\hat{\boldsymbol{\alpha}})$, of which only $\hat{\lambda}_k$ of (7) is used. For the risk measure discussed in this paper, it suffices to compute these estimates for cells $k$ which are sample uniques, that is, $f_k = 1$.

Equating the partial derivative of the function of (6) with respect to $\beta_0$ to zero we obtain $\sum_{k' \in M} \lambda_{k'}(\hat{\boldsymbol{\alpha}}) = \sum_{k' \in M} f_{k'}$, and other derivatives yield moment identities. Note, however, that these desirable identities hold for $\lambda_{k'}(\hat{\boldsymbol{\alpha}})$ which are obtained for a fixed $k = (k_1, k_2)$, and not for our final estimates in (7), which are the ones we use in the sequel.

With the estimate of (7), recalling $\lambda_k = N\gamma_k\pi_k$ and setting $U = \{k : f_k = 1\}$, the set of sample uniques, we now apply the Poisson formulas (3), see also (1), to obtain the risk estimates

$$(8) \quad \hat{\tau}_1 = \sum_{k \in U} e^{-\hat{\lambda}_k(1-\pi_k)/\pi_k}, \quad \hat{\tau}_2 = \sum_{k \in U} \frac{1}{\hat{\lambda}_k(1-\pi_k)/\pi_k}[1 - e^{-\hat{\lambda}_k(1-\pi_k)/\pi_k}].$$

In our experiments we defined neighborhoods $M$ of $k$ by varying around $k$ coordinates corresponding to attributes that are ordinal, and using close values in non-ordinal attributes when possible (e.g., in Occupation). Attributes in which closeness of values cannot be defined remain constant in the whole neighborhood. Thus in our experiments, neighborhoods always consist of individuals of the same Sex. For more details see Section 4.

## 4. Experiments with neighborhoods

We present a few experiments. They are preliminary as already mentioned and more work is needed on the approach itself and on classifying types of data for which it might work.

In the experiments we used our own versions of the Argus and log-linear models methods, programmed on the SAS system. Throughout our experiments two log-linear models are considered, one of independence of all attributes, the other including all two-way interactions.

The weights $w_i$ for the *Argus* method in all our examples were computed by post-stratification on Sex by Age by Geographical location (the latter is not one of the attributes in any of the tables, but it was used for post-stratification). These variables are commonly used for post-stratification, other strata may give different, and perhaps better results.

In all experiments we took a real population data file of size $N$ given in the form of a contingency table with $K$ cells, and from it we took a simple random sample of size $n$. Since the population and the sample are known to us, we can compute the *true values* of $\tau_1$ and $\tau_2$ and their estimates by the different methods, and compare.

**Example 1.** In this small example the population consists of a small extract from the 1995 Israeli Census with individuals of age 15 and over, with $N = 15,035$ and $K = 448$. From this population we took a random sample of size $n = 1,504$, using a fixed sampling fraction, that is $\pi_k = n/N$ for all $k$. The sampling fraction is constant in all our experiments. The attributes (with number of levels in parentheses) were Age Groups (32), and Income Groups (14), both ordinal.

As mentioned above, throughout our experiments two log-linear models are considered, one of independence, the other including all two-way interactions (which in the present case of two attributes, is a saturated model). In this experiment we tried our proposed smoothing polynomial approach of (5) for $t = 1, 2$. We considered one type of neighborhood here, constructed by changing each attribute value in $k$ by at most 3 values up or down, that is, the neighborhood of each cell $k$ is

$$(9) \qquad M = \{k' : \max_{1 \leq i \leq m} |k'_i - k_i| \leq c\},$$

with $m = 2$, $c = 3$ and hence size $|M| = 49$. For cells near the boundaries some of the cells in their neighborhoods do not exist; here we set non-existing cells' frequencies to be zero, but other possibilities can be considered.

TABLE 1

|  | Example 1 | | Example 2 | |
|---|---|---|---|---|
| Model | $\tau_1$ | $\tau_2$ | $\tau_1$ | $\tau_2$ |
| True Values | 2 | 12.4 | 2 | 19.9 |
| Argus | 7.8 | 19.6 | 14.7 | 37.2 |
| Log Linear Model: | | | | |
| Independence | 0.06 | 6.7 | 0.01 | 9.8 |
| Log Linear Model: | | | | |
| 2-Way Interactions | 0.01 | 8.6 | 1.4 | 19.6 |
| Smoothing $t=1$ $|M|=49$ | 3.2 | 12.0 | 7.0 | 22.5 |
| Smoothing $t=2$ $|M|=49$ | 1.7 | 10.4 | 4.8 | 19.0 |

Table 1 presents the true $\tau$ values and their estimates by the methods described above.

**Example 2.** The population consists of an extract from the 1995 Israeli Census, $N = 37,586$, $n = 3,759$, and $K = 896$. The attributes are Sex(2) * Age Groups (32) * Income Groups(14).

We applied the smoothing polynomial of (5) for $t = 1, 2$ and neighborhoods obtained by varying the attributes of Age and Income as in Example 1 and keeping Sex fixed. In other words we used the neighborhoods

$$(10) \qquad M = \{k' : k'_1 = k_1, \max_{2 \leq i \leq m} |k'_i - k_i| \leq c\},$$

with $m = 3$, $c = 3$ which are like (9) on each sub-table of males and females. The results are given in Table 1.

**Example 3.** Population: an extract from the 1995 Israeli Census. $N = 37,586$, $n = 3,759$, $K = 11,648$. Attributes: Sex(2) * Age Groups (32) * Income Groups(14) * Years of Study (13).

We applied the smoothing polynomial of (5) for $t = 2$ and neighborhoods obtained by fixing Sex, so neighborhoods are as in (10), but with $m = 4$, $c = 2$, and since we now vary three variables, each over a range of five values, we have $|M| = 125$. The results are given in Table 2.

TABLE 2

| Model | $\tau_1$ | $\tau_2$ |
|---|---|---|
| True Values | 187 | 452.0 |
| Argus | 137.2 | 346.4 |
| Log Linear Model: | | |
| Independence | 217.3 | 518.0 |
| Log Linear Model: | | |
| 2-Way Interactions | 167.2 | 432.8 |
| Smoothing $t=2$ $|M|=125$ | 170.7 | 447.9 |

Table 3

| Model | $\tau_1$ | $\tau_2$ |
|---|---|---|
| True Values | 191 | 568.0 |
| Argus | 79.2 | 315.6 |
| Log Linear Model: | | |
| Independence | 364.8 | 862.3 |
| Log Linear Model: | | |
| 2-Way Interactions | 182.2 | 546.2 |
| Smoothing $t = 2$ $|M| = 545$ | 139.6 | 509.1 |
| Smoothing $t = 2$ $|M| = 625$ | 154.7 | 528.5 |
| Smoothing $t = 2$ $|M| = 1025$ | 215.7 | 647.2 |

Table 4

| Model | $\tau_1$ | $\tau_2$ |
|---|---|---|
| True Values | 5 | 36.9 |
| Argus | 7.7 | 35.5 |
| Log Linear Model: | | |
| Independence | 6.4 | 44.2 |
| Log Linear Model: | | |
| 2-Way Interactions | 1.1 | 26.4 |
| Smoothing $t = 2$ $|M| = 125$ | 3.3 | 31.3 |

**Example 4.** Population: an extract from the 2001 UK Census File. $N = 944,793$, $n = 18,896$, $K = 152,100$. Attributes: Sex (2) * Age Groups (25) * Number of Persons in Household (9) * Education Qualifications (13) * Occupation (26).

We applied the smoothing polynomial of (5) for $t = 2$ and neighborhoods defined by fixing Sex and varying all other variables, including Occupation, which was coded as ordinal. The neighborhoods are

$$(11) \qquad M = \{k' : k'_1 = k_1, \max_{2 \le i \le m} |k'_i - k_i| \le c, \sum_i |k'_i - k_i| \le d\},$$

with $m = 5$, $c = 2$ and $d = 6, 8$, resulting in neighborhood sizes $|M| = 545$ and 625, respectively. We also tried $c = 3$, $d = 6$ and hence $|M| = 1025$. The results are given in Table 3.

**Example 5.** Population: an extract from the 1995 Israeli Census. $N = 248,983$, $n = 2,490$, $K = 8,800$. Attributes: Sex(2)* Age Groups(16) * Years of Study (25) * Occupation (11).

We applied the smoothing polynomial of (5) for $t = 2$ and neighborhoods obtained by varying three attributes and fixing Sex so neighborhoods as in (10) with $m = 4$, $c = 2$, and $|M| = 125$. The results are given in Table 4.

**Example 6.** Population: an extract from the 1995 Israeli Census. $N = 746,949$, $n = 14,939$, $K = 337,920$. Attributes: Sex (2) * Age Groups (16) * Years of Study (10) * Number of Years in Israel (11) * Income Groups (12) * Number of Persons in Household (8). Note that this is a very sparse table.

We applied the smoothing polynomial of (5) for $t = 2$ and neighborhoods obtained by varying all attributes except for Sex which was fixed. Neighborhoods are as in (11) with $m = 6$, $c = 2$, $d = 4, 6$, and $|M| = 581$ and $1,893$, respectively. The results are given in Table 5.

TABLE 5

| Model | $\tau_1$ | $\tau_2$ |
|---|---|---|
| True Values | 430 | 1,125.8 |
| Argus | 114.5 | 456.0 |
| Log Linear Model: Independence | 773.8 | 1,774.1 |
| Log Linear Model: 2-Way Interactions | 470.0 | 1,178.1 |
| Smoothing $t=2$ $|M|=581$ | 287.1 | 988.4 |
| Smoothing $t=2$ $|M|=1,893$ | 471.1 | 1,240.2 |

TABLE 6

| Model | $\tau_1$ | $\tau_2$ |
|---|---|---|
| True Values | 42 | 171.2 |
| Argus | 20.7 | 95.4 |
| Log Linear Model: Independence | 28.8 | 191.5 |
| Log Linear Model: 2-Way Interactions | 35.8 | 164.1 |
| Smoothing $t=2$ $|M|=545$ | 37.1 | 175.1 |

**Example 7.** Population: an extract from the 1995 Israeli Census. $N = 746,949$, $n = 7,470$, $K = 42,240$. Attributes: Sex (2) * Age Groups (16) * Years of Study (10) * Number of Years in Israel (11) * Income Groups (12).

We applied the smoothing polynomial of (5) for $t = 2$ and neighborhoods obtained by varying all attributes except for Sex which was fixed. Neighborhoods are as in (11) with $m = 5$, $c = 2$, $d = 6$, and $|M| = 545$. Smaller neighborhood did not yield good estimates. The results are given in Table 6.

**Discussion of examples** The log-linear model method was tested in Skinner and Shlomo [11] and references therein, and it seems to yield good results for experiments of the kind done here. Di Consiglio et al. [5] presented experiments for individual risk assessment with Argus, which seems to perform less well than the log-linear method in many of our experiments with global risk measures. Our new method still requires fine-tuning. At present the results seem comparable to the log-linear method, and it seems to be computationally somewhat simpler and faster.

Naturally, more variables and sparse data sets with a large number of cells are typical and need to be tested. Such files will cause difficulties to any method, and this is where the different methods should be compared. In sparse multi-way tables, model selection will be crucial but difficult for the log-linear method, and perhaps simpler for the smoothing approach. We also think that our method may be easier to modify to complex sampling designs.

Our proposed method is at a preliminary stage and requires more work. Particular directions are the following:
**1.** Adjust the estimates $\hat{\gamma}_k$ of (7) to fit known population marginals obtained from prior knowledge and sampling weights. In log-linear models the total sum of these estimates corresponds to the sample size, but as commented in Section 3 this is not the case with the smoothing estimates of (7).
**2.** Use goodness of fit measures and information on population marginals and sampling weights to select the type and size of the neighborhoods, and the degree of the smoothing polynomial in (5). We have observed in experiments that when the sum of all estimates matches the sample size, we obtain good risk measure estimates, and further matching to marginals may improve the estimates.

**3.** Extend the smoothing approach to the more general Negative Binomial model which subsumes both the Poisson model implemented here, and the Negative Binomial discussed in Section 2.

**4.** Apply this method also for individual risk measure estimates, which are important in themselves, and may also shed more light on efficient neighborhood and model selection. Our preliminary experiments suggest that the smoothing approach performs relatively well in estimating individual risk.